\documentclass[pra,amsmath,amssymb,amsfonts,twocolumn,nofootinbib]{revtex4}

\newcommand{\one}{\mbox{$1 \hspace{-1.0mm}  {\bf l}$}}

\usepackage{amsmath}
\usepackage{amssymb}
\usepackage{exscale}
\usepackage{graphicx}
\usepackage{epsfig}

\newtheorem{theorem}{Theorem}
\newtheorem{lemma}[theorem]{Lemma}

\newenvironment{proof}{\vspace{1.5ex}\par\noindent\textbf{Proof}}%
    {\hspace*{\fill}$\Box$\vspace{1.5ex}\par}

\begin{document}

\title{Quantum simulations under translational symmetry}

\author{Christina V. Kraus, Michael M. Wolf, J. Ignacio Cirac}
\affiliation{ Max-Planck-Institute for Quantum Optics,
 Hans-Kopfermann-Str.\ 1, D-85748 Garching, Germany.}

\begin{abstract}
We investigate the power of quantum systems for the simulation of
Hamiltonian time evolutions on a cubic lattice under the constraint
of translational invariance. Given a set of translationally
invariant local Hamiltonians and short range interactions we
determine time evolutions which can and those that can not be
simulated. Whereas for general spin systems no finite universal set
of generating interactions is shown to exist, universality turns out
to be generic for quadratic bosonic and fermionic nearest-neighbor
interactions when supplemented by all translationally invariant
on-site Hamiltonians.

\end{abstract}

\date{\today}

\maketitle

\section{Introduction}

One of the most promising applications of quantum computers is
that of being a working horse for physicists who want to determine
the time evolution of a theoretically modeled quantum system. On a
classical computer this is a daunting task burdened by the
notorious exponential growth of the underlying Hilbert space. As
pointed out by Feynman \cite{Feynman,Lloyd}, quantum systems can,
however, be used to efficiently simulate quantum mechanical time
evolutions---provided that we have sufficient coherent control on
the system. In this direction enormous progress has been made
during the last years, in particular in systems of optical
lattices \cite{opticallattices} and ion traps
\cite{iontraps,Diego}. Moreover, it was realized that quantum
simulators \cite{Jane} are much less demanding than quantum
computers and, in fact, pioneering experiments simulating quantum
phase transitions in systems of cold atomic gases
\cite{opticallattices} have already turned some of the visions
\cite{QPTheorie} into reality.

One of the  fundamental questions in the field of quantum
simulations is the following: Given a set of interactions we can
engineer with a particular system, which are the Hamiltonians that
can be simulated? Concerning \emph{gates}, i.e., discrete time
unitary evolutions, it has been shown in the early days of quantum
information theory that almost any two-qubit gate is
\emph{universal} \cite{2universality}. Similarly,  any fixed
entangling two-body interaction was shown to be capable of
simulating any other two-body Hamiltonian when supplemented by the
set of all local unitaries \cite{2body}. The many-body analogue of
 this problem was solved in
 \cite{manybody} and the efficiency of quantum simulations was
 studied in various contexts (cf. \cite{eff}).

All these schemes are based on the addressing of sites, i.e., local
control. Imagine now that we have a chain in which we cannot address
each particle individually but only apply global single-particle and
nearest-neighbor interactions. Can we simulate the evolution of a
next-to-nearest neighbor interaction Hamiltonian, or obtain some
long-range (e.g., dipole) coupling, or even a three-particle
interaction Hamiltonian?

In this article we  will concentrate on the case in which the
interactions at hand are short range and  \emph{translationally
invariant} as it is (approximately) the case in different
experimental set-ups, like in the case of atoms in optical
lattices or in many other systems that naturally appear in the
context of condensed matter and statistical physics. In order to
make the problem mathematically tractable and to exploit its
symmetries we will consider periodic boundary conditions, even
though typically physical systems have open ones. In this sense,
our results may not be directly applicable to certain physical
situations. In any case, we expect that our work will be a step
forward to the establishment of what can and cannot be simulated
with certain quantum systems. We will consider three different
systems: spins, fermions and bosons. A summary of the results of
this work is given in the following section.

\section{statement of the problem and summary of results}

Consider a cubic lattice of $N$ sites with periodic boundary
conditions in arbitrary spatial dimension. Assume that we can
implement every Hamiltonian from a given set
$\mathcal{S}=\{\mathcal{H}_1,  \mathcal{H}_2, \ldots \}$ of
translationally invariant Hamiltonians and in this way achieve every
unitary time evolution of the form $e^{i \mathcal{H}_j t}$ for
arbitrary $ t\in \mathbb{R}$. Note that this assumes that both
$\pm\mathcal{H}_j$ are available. The question we are going to
address is, which evolutions can be simulated by concatenating
evolutions generated by the elements of $\mathcal{S}$. Our main
interest lies in sets which contain all on-site Hamiltonians and
specific nearest-neighbor interactions.

The natural language for tackling this problem is the one of Lie
algebras \cite{FultonHarris, Cornwell} since the set of reachable
interactions is given by the Lie algebra $\mathcal{L}$ generated by
the set $i\mathcal S$. This follows from the \textit{Lie -Trotter
formulae} \cite{LieTrotter}
\begin{eqnarray}
e^{\alpha L_k + \beta L_l}&=&\lim_{n \rightarrow
\infty}\left(e^{\alpha
L_k/n}e^{\beta L_l/n} \right)^n,\quad\alpha,\beta\in\mathbb{R},\nonumber \\
e^{[L_k,L_l]}&=&\lim_{n \rightarrow
\infty}\left(e^{L_k/\sqrt{n}}e^{L_l/\sqrt{n}}e^{-L_k/\sqrt{n}}e^{-L_l/\sqrt{n}}
\right)^n,\nonumber
\end{eqnarray}
where $L_k$ is a representation of the generator $i\mathcal{H}_k$.
When applying the Lie-Trotter formulae to the elements of
$i\mathcal{S}$ we can obtain all commutators and real linear
combinations of its elements, i.e., we end up with the Lie algebra
generated by $\mathcal{S}$. Conversely, it follows from the
Baker-Campbell-Haussdorff formula \cite{BCH} that all simulatable
interactions can be written in this way. We will study the cases of
$D$-dimensional `spin' systems ($\mathcal{L}\subset su_{D^N}$) as
well as quadratic Hamiltonians in fermionic ($\mathcal{L}\subset
so_{2N}$) and bosonic ($\mathcal{L}\subset sp_{2N}$) operators. The
following gives a simplified summary of the main results.
Hamiltonians and interactions are meant to be translationally
invariant throughout and it is assumed that interactions along
different directions can be implemented
independently.\begin{itemize}
    \item \emph{Fermions:} All simulated evolutions have
     \emph{real} tunneling/hopping amplitudes. Within this set generic
     nearest-neighbor interactions are universal for the simulation
    of any translationally invariant interaction when supplemented with all on-site
    Hamiltonians. Whereas for cubes with odd edge length the proof
    of universality requires interactions along all axes and
    diagonals, the diagonals are not required for even
    edge length.

    \item \emph{Bosons:} All simulated evolutions are point
    symmetric. Within this set every
     nearest-neighbor interaction (available along axes and diagonals) is universal for the simulation
    of any translationally invariant interaction when supplemented with all on-site
    Hamiltonians.
    \item \emph{Spins:} There is no universal set of nearest-neighbor interactions. Moreover, if $f$ is a factor of the edge length $m$
    of the cubic lattice
    then there is no universal set of interactions with
    interaction range smaller than $f$. In particular, if $m$ is even, not all
    next-to-nearest neighbor interactions can be simulated from
    nearest neighbor ones. Sets of Hamiltonians that can
    be simulated are constructed.
\end{itemize}

Whereas in the case of quadratic bosonic and fermionic
Hamiltonians a rather exhaustive characterization of simulatable
time evolutions is possible, a full characterization of
simulatable spins systems still remains an open problem.

We will start with introducing some preliminaries  on quadratic
Hamiltonians in Sec.\ref{Sec:quadraticHs}. Sec.\ref{Fermi} will
then treat fermionic and Sec.\ref{Bosonic} bosonic systems. Both
start with the one-dimensional case which is then generalized to
arbitrary dimensional cubic lattices. Finally spin systems are
addressed in Sec.\ref{Spin}.
\section{Quadratic Hamiltonians}\label{Sec:quadraticHs}

This section will introduce the basic notions and the notation used
in Secs. \ref{Fermi},\ref{Bosonic}. The presentation is a collection
of tools widely used in the literature on translationally invariant
quasi-free fermionic \cite{LSM, Araki, Bravyi, Wolf} and bosonic
\cite{Audenaert-Eisert-Werner, Schuch} systems. We consider a system
of $N$ fermionic or bosonic modes characterized by a quadratic
Hamiltonian
\begin{equation}\label{H_def}
\mathcal{H}=\sum_{k,l=1}^N A_{kl}a_k
a_l+B_{kl}a_ka_l^{\dagger}+C_{kl}a_k^{\dagger}
a_l+D_{kl}a_k^{\dagger}a_l^{\dagger}.
\end{equation}
Here, $a_k^{\dagger}$ and $a_k$ are creation and annihilation
operators satisfying the canonical (anti)-commutation relations
\begin{eqnarray}
\mbox{CAR: } \{a_k, a_l\}&=&0,\;\;\{a_k, a_l^{\dagger}\}=\delta_{kl}\;\;\mbox{(fermions)},\\
\mbox{CCR: }\left[a_k, a_l\right]&=&0,\;\;[a_k,
a_l^{\dagger}]=\delta_{kl}\;\;\mbox{(bosons)}.
\end{eqnarray}
By defining a vector $\alpha=(a_1,\ldots,a_N,a_1^\dagger,\ldots
a_N^\dagger)$ and a \emph{Hamiltonian matrix}
\begin{equation}
\tilde{H}=\left(%
\begin{array}{cc}
  A & B \\
  C & D \\
\end{array}%
\right)
\end{equation}
 Eq.\eqref{H_def} can be written in the compact form $
\mathcal{H}=\alpha \tilde{H} \alpha^T$. The Hermiticity of
$\mathcal{H}$ implies the relations
\begin{equation}
B=B^\dagger,\;C=C^\dagger\;,A=D^\dagger.
\end{equation}
We will identify Hamiltonians which differ by multiples of the
identity as they give rise to undistinguishable time evolutions. The
commutation relations can then be exploited to symmetrize the
Hamiltonian matrix $\tilde{H}$ such that
\begin{equation}
A=\tau A^T, D=\tau D^T, B=\tau C^T,
\end{equation}
where $\tau=1$ for bosons and $\tau=-1$ in the case of fermions.
Instead of working with $2N$ creation and annihilation operators it
is often convenient to introduce $2N$ hermitian operators $c_k$ via
\begin{equation}
c_k=(a_k^{\dagger}+a_k)/\sqrt{2},\;\;c_{k+N}=i(a_k^{\dagger}-a_k)/\sqrt{2}.
\end{equation}
In the case of fermions these are the Majorana operators obeying the
anti-commutation relation
\begin{equation}
\{c_k,c_l\}=\delta_{kl}.
\end{equation}
For bosons the $c_k$ are the position  and $c_{k+N}$ momentum
operators, and the commutation relations can be expressed in terms
of the symplectic matrix $\sigma$ via
\begin{equation}\label{CCR}
[c_k,c_l]=i\sigma_{kl},\;\;\sigma =\left(
          \begin{array}{cc}
            0 & \one \\
            -\one & 0 \\
          \end{array}
        \right), \;\; \one\in \mathbb{R}^{N \times N}.
\end{equation}
Eq. \eqref{H_def} can now be written in the form
\begin{equation}\label{H_c}
\mathcal{H}=\frac{\sqrt{\tau}}{2}\sum_{k,l}H_{kl}c_kc_l,\ \ \
H=\left(
          \begin{array}{cc}
            X & W \\
            \tau W^T & Y \\
          \end{array}
        \right).
\end{equation}
Exploiting again the commutation relations we can choose the
Hamiltonian matrix $H$ real and (anti-) symmetric  with $H=\tau
H^T$. The Hamiltonian matrices of the two representations are
related via
\begin{equation*}
\tilde{H}= \frac{\sqrt{\tau}}4 {\small\left(%
\begin{array}{cc}
  X-Y-i(W+\tau W^T) & X+Y+i(W-\tau W^T) \\
  X+Y-i(W-\tau W^T) & X-Y+i(W+\tau W^T) \\
\end{array}%
\right) }.\vspace*{3pt}
\end{equation*}

{\it Time-evolution:} We are interested in time-evolutions
generated by quadratic Hamiltonians of the form in Eq.
\eqref{H_def}. These are canonical transformations which preserve
the (anti-) commutation relations and act (in the Heisenberg
picture) linearly on the $c_k$'s:
\begin{equation}\label{canonical transformation}
e^{i\mathcal{H}t}c_ke^{-i\mathcal{H}t} = \sum_{l=1}^N T_{lk}
c_l\;.
\end{equation}
In the fermionic case the CAR are preserved iff $T \in O(2N)$ is an
element of the orthogonal group in $2 N$ dimensions. This group has
two components corresponding to elements with determinant $\pm 1$.
As time evolution has to be in the part connected to the identity
(for $t=0$) we have that $T \in SO(2N)$ is an element of the special
orthogonal group. For bosons the preservation of the commutation
relations implies that $T$ is a symplectic matrix, i.e. $T\sigma
T^T=\sigma$. Both groups $SO(2N)$ and $Sp(2N)$ are Lie groups and we
can express $T$ in terms of the exponential map acting on the
respective Lie algebra, i.e., $T=e^{tL}$. From the infinitesimal
version of Eq.(\ref{canonical transformation}) we obtain a simple
relation between the generator $L$ and the Hamiltonian matrix $H$:
\begin{eqnarray}
L&=&-H\;\;\,\,\, \mbox{for fermions}\\
L&=&H\sigma^T\;\; \mbox{for bosons}\label{L-Boson}.\vspace*{3pt}
\end{eqnarray}

{\it Translational invariant systems:} We will throughout consider
translationally invariant systems on cubic lattices in $d$ spatial
dimensions with periodic boundary conditions. Hence, the indices of
the Hamiltonian matrix $H_{kl}$
 which correspond to two points on the lattice  are
$d$-component vectors $k,l\in\mathbb{Z}^d_m$ where $m$ is the edge
length of the cube, i.e., $N=m^d$. The translational invariance is
expressed by the fact that the matrix elements $G_{kl}$, of the
blocks $G \in \{X,Y,W\}$ of $H$ depend only on the relative distance
$k-l$. Taking into account the periodic boundary conditions, $k-l$
is understood modulo $m$ in each component. Such matrices are called
\textit{circulant}, and we will denote by $\mathcal{C}_A$ and
$\mathcal{C}_S$ the set of circulant symmetric and antisymmetric
matrices, respectively. All circulant matrices can be diagonalized
simultaneously by Fourier transformation
\begin{eqnarray}\hat{G}\equiv  \mathcal{F}^{\otimes d}G \mathcal{F}^{\dagger \otimes
d}&=& \mbox{diag}\left[\sum_{k \in \mathbb{Z}^d_m} G_k e^{-\frac{2
\pi i }{m}kl} \right]_l,\label{FT}\\
\mathcal{F}_{pq}&=&\frac{1}{\sqrt{m}}e^{\frac{2 \pi i}{m}pq},\;\;
p,q \in \mathbb{Z}_m,
\end{eqnarray}
where $G_k\equiv G_{k,0}$ is the entry of the $k$-th off-diagonal
of the matrix $G$. It follows from \eqref{FT} that all circulant
matrices mutually commute.

\section{Simulations in fermionic systems}\label{Fermi} In this section we study the set of
 interactions that can be simulated in a
translationally invariant fermionic system starting with quadratic
local transformations and nearest neighbor-interactions. Making use
of the fact that the blocks $X,Y$ and $W$  in Eq.(\ref{H_c})
mutually commute we calculate the commutator $L^{\prime
\prime}=[L,L^{\prime}]$ of two generators $L$ and $L^{\prime}$ given
by \begin{equation}\label{L'} L^{(\prime)}=\left(
      \begin{array}{cc}
        X^{(\prime)} & W^{(\prime)} \\
        -W^{(\prime)T} & Y^{(\prime)} \\
      \end{array}
    \right)
\end{equation}
and obtain 
\begin{equation}\label{Aprime-Fermi}
L^{\prime \prime}={\small \left(
                    \begin{array}{cc}
                      X^{\prime \prime}& W^{\prime \prime} \\
                      -W^{\prime \prime T} & -X^{\prime \prime} \\
                    \end{array}
                  \right)},\; \begin{array}{lcl}
                             X^{\prime \prime}&=&W^{\prime}W^T-WW^{\prime T} \\
                             W^{\prime
                             \prime}&=&W(Y^{\prime}-X^{\prime})\\
                             &&-W^{\prime}(Y-X)
                           \end{array}\hspace{-9pt}
\end{equation}
Note that by Eq.(\ref{Aprime-Fermi}) every commutator has the
symmetry $Y''=-X''$. Hence, if we start with a set $\mathcal S$ of
Hamiltonians with corresponding generators
$\mathcal{S_L}=\{L_1,L_2,\ldots\}$, then every element of the
generated Lie algebra $\mathcal{L}$ has this form up to linear
combinations of elements in $\mathcal{S_L}$. On the level of
Hamiltonians this symmetry corresponds to real tunneling/hopping
coefficients $B_{kl}=-C_{lk}\in\mathbb{R}$ in Eq.(\ref{H_def}). We
will denote by $\mathcal{R}$ the vector space of all matrices of the
form \eqref{L'} for which $Y=-X$.

Let us now introduce the elements of the set $\mathcal{S_L}$
corresponding to all local Hamiltonians and specific
nearest-neighbor interactions. Every generator $L$ of a local
Hamiltonian is proportional to
\begin{equation}\label{E}
E=\left(
                  \begin{array}{cc}
                    0 & \one \\
                    -\one & 0 \\
                  \end{array}
                \right).
\end{equation}
For giving an explicit form to the nearest-neighbor interaction,
we define a matrix $M^{(v)}$ via
\begin{equation}\label{M}
M^{(v)}_{kl}=\delta_{l,k+v},
\end{equation}
where $v,k,l \in \mathbb{Z}_m^d$ and the addition is modulo $m$ in
each of the components. This leads to the properties
\begin{equation}\label{M^m=1}
M^{(v_1)}M^{(v_2)}=M^{(v_1+v_2)},\ \ M^{(0)}=\one.
\end{equation}
Moreover, we define the matrices
\begin{eqnarray}\label{Blocks-Defined-Fermi}
M^{(v)}_+&=&M^{(v)}+M^{(-v)},\ \
M^{(v)}_-=M^{(v)}-M^{(-v)},\ \ \ \ \label{HX-HW-defined} \\
H_X^{(v)}&=&{\small \left(
            \begin{array}{cc}
               M^{(v)}_- & 0 \\
              0 & -M^{(v)}_- \\
            \end{array}
          \right)},\label{HX}\\
          H^{(v)}_{W(\pm)}&=&{\small\left(
            \begin{array}{cc}
              0 &  M^{(v)}_{(\pm)} \\
               M^{(v)T}_{(\pm)} &0 \\
            \end{array}
          \right)},\label{HW}
\end{eqnarray}
where the indices $X$ and $W$ refer to a non-vanishing $X$- and
$W$-block respectively. Denoting by $e_i \in \mathbb{Z}_m^d$ the
basis vectors  $(e_i)_j = \delta_{ij}$, every Hamiltonian matrix
corresponding to a nearest-neighbor interaction along $e_i$ is of
the form
\begin{eqnarray}\label{XYZ_NN}
&&H_0\equiv \left(%
\begin{array}{cc}
  X_0 & W_0 \\
  -W_0^T & Y_0 \\
\end{array}%
\right)=\\
&&\left(%
\begin{array}{cc}
  x M^{(e_i)}_- & w M^{(e_i)}+\tilde{w} M^{(-e_i)} \\
  -(\tilde{w} M^{(e_i)}+w M^{(-e_i)}) & y M^{(e_i)}_- \\
\end{array}%
\right),\nonumber
\end{eqnarray}
where $x, y, w, \tilde{w} \in \mathbb{R}$. We will now start
studying one-dimensional systems and then generalize to the
$d$-dimensional case.

\subsection{Simulations in one-dimensional fermionic systems} In
this section we consider quadratic fermionic Hamiltonians with
translational symmetry on a ring of $m$ sites. We will give an
exhaustive characterization of nearest-neighbor Hamiltonians which
are universal for the simulation of all interactions obeying the
symmetry $Y=-X$, when supplemented by all on-site Hamiltonians.
The results depend on whether $m$ is even or odd.

\begin{theorem}\label{Simulation-Fermi-1d}
Consider a translationally invariant fermionic systems of $m$ sites
on a ring with periodic boundary conditions. Starting with all
one-particle transformations which are proportional to the matrix
$E$ defined in \eqref{E} and one nearest-neighbor interaction $H_0$
of the form \eqref{XYZ_NN} we can simulate the following set of
interactions depending on the symmetry properties of $X_0, Y_0$ and
$W_0$:
\begin{enumerate}
\item $m=2n+1$ odd:
\begin{enumerate}
\item $X_0=Y_0, W_0 \in \mathcal{C}_S$: No further interaction can be
simulated.\label{odd-a}
\item $X_0 \neq Y_0$ or $W_0 \notin \mathcal{C}_S$: The space $\mathcal{R}$ (i.e. $X=-Y$) can be simulated.\label{odd-b}
\end{enumerate}
\item $m=2n$ even:
\begin{enumerate}
\item $X_0=Y_0, W_0 \in \mathcal{C}_S$: No further interaction can be
simulated.\label{even-a}
\item $W_0\in \mathcal{C}_A$ or $W_0=0$: The space spanned by
$$\mathcal{I}\equiv\{H^{((2k-1)e_1)}_{X},H^{(2k
e_1)}_{W+},H^{((2k-1)e_1)}_{W-}\}_{k\in\mathbb{N}}$$ can be
        simulated. (For the definition of $H_X$ and $H_W$ see Eqs. \eqref{HX}, \eqref{HW} ).\label{even-b}
\item  $W_0 \notin \mathcal{C}_A, H_0 \in \mathcal{R}$: The space $\mathcal{R}$ can be
    simulated.\label{even-c}
\item $W_0 \notin \mathcal{C}_A, H_0 \notin \mathcal{R}$:
The space spanned by $$\mathcal{J}\equiv\{H^{(ke_1)}_{X}, H^{(2k
e_1)}_{W}, H^{((2k-1) e_1)}_{W}-H^{((2(k-1)-1) e_1)}_{W}
\}_{k\in\mathbb{N}}$$ can be simulated.\label{even-d}
\end{enumerate}
\end{enumerate}
\end{theorem}

\begin{proof}
For the proof we will need the relations
\begin{eqnarray}
[H_X^{(ke_1)},E]&=&2H_{W-}^{(ke_1)},\label{HX-E}\\
\,[E,H_W^{(ke_1)}]&=&H_X^{(ke_1)},\label{E-HW}\\ \,
[H_X^{(ke_1)},H_X^{(le_1)}]&=&0,\label{HX-HX}\ \ \ \ \ \\
\,[H_X^{(ke_1)},H_W^{(le_1)}]&=&2(H_W^{((l+k)e_1)}-H_W^{((l-k)e_1)}), \quad\label{HX-HW}\\
\,[H_W^{(ke_1)},H_W^{(le_1)}]&=&H_X^{((l-k)e_1)}\label{HW-HW}.
\end{eqnarray}
For $X_0=Y_0, W_0 \in \mathcal{C}_S$, $[H_0,E]=0$ according to
\eqref{Aprime-Fermi} so that we cannot simulate any
further interaction (up to multiples of $E$). This proves (\ref{odd-a}) and (\ref{even-a}).\\
If $W_0 \notin \mathcal{C}_S$ or $X_0 \neq Y_0$, we will show in the
first step by induction over $k$  that the set $\mathcal{I}$ defined
in (\ref{even-b}) can be simulated. For $k=1$, we can get
$H_X^{(e_1)}$ and $H_{W-}^{(e_1)}$ by taking the commutator of $H_0$
with the one-particle transformation $E$: If $W_0 \in
\mathcal{C}_S$, then $[H_0,E] \sim H_{W-}^{(e_1)}$ and $H_X^{(e_1)}$
can be obtained using \eqref{E-HW}. If $X_0=Y_0$, then $[H_0,E] \sim
H_X^{(e_1)}$ and we get $H_{W-}^{(e_1)}$ by \eqref{HX-E}. If $W_0
\notin \mathcal{C}_S$ and $X_0 \neq Y_0$, then
$$[H_0,E]/(\tilde{w}_0-w)+[[H_0,E],E]/2(x_0-y_0)\sim
H_{W-}^{(e_1)}$$ and according to \eqref{E-HW} we also get
$H_X^{(e_1)}$. From \eqref{HX-HW} we see that we get
$$H_{W+}^{(2e_1)}=[H_X^{(e_1)}, H_{W-}^{(e_1)}]/2+2E.$$
Now let $k\geq 1$. Using \eqref{HX-HW} and \eqref{E-HW} we get
$$[H_X^{(e_1)},H_{W+}^{(2ke_1)}]+H_{W-}^{((2k-1)e_1)}=H_{W-}^{((2(k+1)-1)e_1)}$$
which implies that we also get $H_{X}^{((2(k+1)-1)e_1)}$. As
$$[H_X^{(e_1)},H_{W-}^{((2(k+1)-1)e_1)}]+H_{W+}^{(2e_1)}=H_{W+}^{(2(k+1)e_1)},$$
we have shown that we can simulate $\mathcal{I}$. Using the
relations \eqref{HX-E} - \eqref{HW-HW}, we see that $\mathcal{I}$ is
closed under the commutator bracket.\\
If $m=2n+1$, $\mathcal{I}$ is a basis of all possible interactions
of the space $\mathcal{R}$ because of the periodic boundary
conditions. To see this, define for an arbitrary $k$ the number
$k^{\prime}=k+n+1$. As $2k^{\prime} \mbox{mod}\, m =2k+1$, we see
that
$H_X^{((2k^\prime-1)e_1)}=H_X^{(2ke_1)},H_{W-}^{((2k^\prime-1)e_1)}=H_{W-}^{(2ke_1)}$
and
$H_{W+}^{(2k^\prime e_1)}=H_{W+}^{(2(k+1)e_1)}$, which proves (\ref{odd-b}).\\
Now let $m=2n$. If $W_0 \in \mathcal{C}_A$ or $W_0=0$, then
$$H_0=\tilde{H}+w_0 H_W^{(e_1)},\;\;\;\tilde{H}={\footnotesize \left(
                        \begin{array}{cc}
                          x_0M_-^{(e_1)} & 0 \\
                          0 &  y_0M_-^{(e_1)}\\
                        \end{array}
                      \right)}\notin \mathcal{I}.$$
The elements of $\mathcal{I}$ are the only ones that can be
simulated as
$$[\tilde{H}, H_{W\pm}^{(le_1)}]=(x_0-y_0)[H_X^{(e_1)},
H_{W\pm}^{(le_1)}]\in \mathcal{I}$$
where $l=2k, 2k-1$ respectively and $\tilde{H}$ commutes with $H_{X}^{((2k-1)e_1)}$. This proves (\ref{even-b}).\\
If $W_0 \notin \mathcal{C}_A$ and $H_0 \in \mathcal{R}$, then
$$H_0=x_0H_X^{(e_1)}-\tilde{w}_0H_{W-}^{(e_1)}+(w_0+\tilde{w}_0)H_W^{(e_1)}$$
so that we can extract $H_W^{(e_1)}$. According to \eqref{HX-HW}
$$[H_X^{((2k-1)e_1)},H_W^{(e_1)}]=2(H_W^{(2ke_1)}-H_W^{(-2(k-1)e_1)}),$$
so that we can get $H_W^{(2ke_1)}$ as $H_W^{(0)}=E$, and we can
simulate $H_X^{(2ke_1)}$ using \eqref{E-HW}. It remains to show that
we can simulate $H_W^{((2k-1)e_1)}$. Note that the possibility of
simulating $H_W^{(ke_1)}$ implies that we can get $H_W^{(-ke_1)}$,
as
$$H_W^{(ke_1)}+[[H_W^{(ke_1)},E],E]/2=H_W^{(-ke_1)}.$$
According to \eqref{HW-HW}
$$[H_W^{(e_1)},H_{W+}^{(2ke_1)}]=H_W^{((2k+1)e_1)}+H_W^{(-(2k-1)e_1)}$$
so that we can get $H_W^{((2k-1)e_1)}$ as $H_W^{(-e_1)}$  is available. This proves (\ref{even-c}).\\
Finally we consider the case where $W_0 \notin \mathcal{C}_A$ and
$H_0 \notin \mathcal{R}$. Then
\begin{eqnarray*}
H_0&=&\tilde{H}+\tilde{w_0}H_{W-}^{(e_1)},\\
\tilde{H}&=&{\footnotesize \left(
\begin{array}{cc}
                                                               X_0 & (w_0+\tilde{w}_0)M^{(e_1)}) \\
                                                               -(w_0+\tilde{w}_0)M^{(-e_1)} & Y_0 \\
                                                             \end{array}
                                                           \right)}\notin
                                                           \mathcal{I}.
                                                           \end{eqnarray*}
We will now calculate the commutator of $\tilde{H}$ with all
elements of $\mathcal{I}$ in order to see if we get additional
interactions. From
$$[\tilde{H},H_X^{((2k-1)e_1)}]=-2(w_0 +
\tilde{w}_0)H_{W-}^{(2ke_1)}$$ we see that we can get
$H_{W}^{(2ke_1)}$, as $H_{W+}^{(2ke_1)}\in \mathcal{I}$ and using
\eqref{E-HW} we get $H_X^{(2ke_1)}$. As
$$[H_X^{(ke_1)},H_W^{(2le_1)}]=2(H_W^{((2l+k)e_1)}-H_W^{((2l-k)e_1)}),$$
we have shown that the set $\mathcal{J}$ can be simulated. Using
\eqref{HX-E}-\eqref{HW-HW}, we see that $\{\mathcal{J}, \tilde{H}\}$
is closed under the commutator bracket, which proves (\ref{even-d}).
\end{proof}

\subsection{Simulations in $d$-dimensional fermionic systems}
This section will generalize the previous results to systems in $d$
spatial dimensions. The following theorem shows that certain
 nearest-neighbor
interactions are universal for simulating the space $\mathcal{R}$
(i.e. $Y=-X$) on a $d$-dimensional cube.

\begin{theorem}\label{Fermi-d-dim-odd}
Consider a fermionic systems on a $d$-dimensional translationally
invariant cubic lattice with $m^d$ sites and periodic boundary
conditions. Then the following sets of nearest-neighbor interactions
together with all on-site transformations are complete for
simulating the space $\mathcal{R}$:
\begin{enumerate}
  \item $m=2n+1$ odd: \begin{equation*}\label{H_0Fermi-d-odd}
                H_0^{(e_i)}={\small \left(
                                        \begin{array}{cc}
                                          x_iM_-^{(e_i)} & w_iM^{(e_i)}+\tilde{w}_iM^{(-e_i)} \\
                                          \tilde{w}_iM^{(e_i)}+w_iM^{(-e_i)} & y_iM_-^{(e_i)} \\
                                        \end{array}
                                      \right)},
\end{equation*}
where $i=1, \ldots, d$, $x_i \neq y_i$ or $w_i \neq \tilde{w}_i$ for
all $i$.
  \item $m=2n$ even: $d$ interactions $H_0^{(e_i)}$ of the above form
  where $x_i=-y_i, w_i \neq -\tilde{w}_i$ for all $i$ and $2^d$ interactions of the form
  $H_W^{(\sum_{i=1}^d c_i e_i)}$, $c_i\in \{0,\pm 1\}$.
\end{enumerate}
\end{theorem}

\begin{figure}[tbh]
\begin{center}
\includegraphics[scale=0.7]{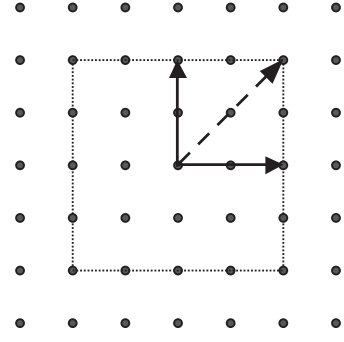}
\end{center} \caption{Two-dimensional fermionic lattice with box $B_2$ (see proof of Thm. \ref{Fermi-d-dim-odd}). The interactions along $e_1$ and
$e_2$ (solid arrows) are used to simulate the interactions in the
direction $2 (e_1 + e_2)$ (dashed arrow).  \label{Graphik}}
\end{figure}

\begin{figure}[t]
\begin{center}
\includegraphics[scale=0.7]{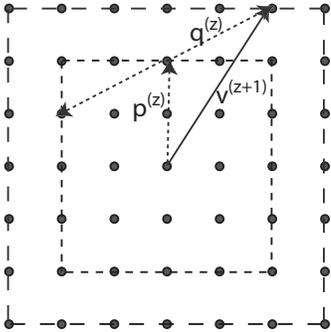}
\end{center} \caption{Boxes $B_z$ and $B_{z+1}$. Any vector $v^{(z+1)} \in B_{z+1}$ can be decomposed
into a sum of vectors from $B_{z}$, i.e.,
$v^{(z+1)}=p^{(z)}+q^{(z)}$ where $p^{(z)},q^{(z)}\in B_z$, such
that $q^{(z)}-p^{(z)}\in B_{z}$ (see proofs of Thms.
\ref{Fermi-d-dim-odd} and \ref{Bose-dd-basis}). \label{Graphik2}}
\end{figure}

\begin{proof}
We start with an odd number of fermions, $m=2n+1$. For the proof we
will consider interactions with a maximal interaction range in each
direction $e_i$ of the lattice. To do so, we define for every
integer $z \in \mathbb{N}$ the $d$-dimensional cube of edge length
$2z$, $B_z=\{v \in \mathbb{Z}_m^d|\, \|v\|_{\infty}\leq z \}$. Then
a Hamiltonian $H^{(v^{(z)})}$ where $v^{(z)}=(v^{(z)}_1, \ldots,
v^{(z)}_d)\in B_z$ couples a given lattice site $s$ only with sites
which lie in a cube of edge size $2 z$ with $s$ in its center. We
will show by induction over $z$ that $H_X^{(v)}$ and $H_W^{(v)}$ can
be simulated. We start with a minimal edge length of 2, i.e. $z=1$
and define $\mathcal{N}_{v}=\{i \,|\, |v_i|=1\}$ with cardinality
$|\mathcal{N}_{v}|$. We will show that $H_X^{(v^{(1)})},
H_W^{(v^{(1)})}$ can be simulated for $|\mathcal{N}_{v^{(1)}}|=1,
\ldots, d$, i.e., for an arbitrary number of non-vanishing
components of the vector $v^{(1)}$. For $|\mathcal{N}_{v^{(1)}}|=1$,
the vector $v^{(1)}$ has only one non-vanishing component $v^{(1)}=
\pm e_i$, and the situation is as the one of theorem
\ref{Simulation-Fermi-1d}. Hence $H_X^{(\pm e_i)}$ and $H_W^{(\pm
e_i)}$ can be simulated for arbitrary $i$. Now let
$|\mathcal{N}_{v^{(1)}}|=r>1$, $j\in \mathcal{N}_{v^{(1)}}$, i.e. we
want to simulate an interaction in the direction of the diagonals as
depicted in figure \ref{Graphik}. As
$|\mathcal{N}_{v^{(1)}-v^{(1)}_je_j}|=r-1$ we know by induction over
the cardinality of $\mathcal{N}_{v^{(1)}}$ that
$H_W^{(v^{(1)}-v^{(1)}_je_j)}$ can be simulated. Then we get
$H_X^{(v^{(1)})}$ as
$$[H_W^{(-v^{(1)}_je_j)},
H_W^{(v^{(1)}-v^{(1)}_je_j)}]=H_X^{(v^{(1)})},$$ and
$H_W^{(v^{(1)})}$ can be obtained according to theorem
\ref{Simulation-Fermi-1d}. Now we consider boxes with edge length
bigger than 2 assuming that $H_X^{(v^{(z)})}$ and $H_W^{(v^{(z)})}$
can be constructed for all $v^{(z)} \in B_z$, and let $v^{(z+1)}=
(v_1^{(z+1)}, \ldots, v_1^{(z+1)}) \in B_{z+1}$. First we show that
there exist $p^{(z)}, q^{(z)} \in B_z$ such that
$p^{(z)}+q^{(z)}=v^{(z+1)}, p^{(z)}-q^{(z)}\in B_z$ (see figure
\ref{Graphik2}). Therefore we define the set
$\mathcal{Z}_v=\{i||v_i^{(z+1)}|=z+1\}$. If we take $q^{(z)}=\sum_{i
\in \mathcal{Z}_v}\frac{v_i}{|v_i|}e_i$ and
$p^{(z)}=v^{(z+1)}-q^{(z)}$ then by definition $p^{(z)}, q^{(z)} \in
B_z$ and $p^{(z)}+q^{(z)}=v^{(z+1)}$. As for all $i\in
\mathcal{Z}_v$ we have $|(p^{(z)}-q^{(z)})_i|=(|v_i|-2)\leqq z-1$
and for all $i\notin \mathcal{Z}_v$ we have
$|(p^{(z)}-q^{(z)})_i|=(|v_i|)\leqq z$, it follows that
$p^{(z)}-q^{(z)}\in B_z$. Using now the commutator relation
$$[H_X^{(q^{(z)})},H_W^{(p^{(z)})}]=2(H_W^{(v^{(z+1)})}-H_W^{(p^{(z)}-q^{(z)})}),$$
(see Eq.\eqref{HX-HW}) we obtain that $H_W^{(v^{(z+1)})}$ can be
simulated, and from \eqref{E-HW} we know
that we also get $H_X^{(v^{(z+1)})}$.\\
Now consider the case where $m=2n$. If $x_i=-y_i, w_0 \neq
-\tilde{w}_i \; \forall i$, we can simulate $H_X^{(k e_i)}$ and
$H_X^{(k e_i)}$ for all $i$ and $k$ according to theorem
\ref{Simulation-Fermi-1d} (\ref{even-c}). Like in the case $m=2n+1$
we can simulate $H_X^{(kv^{(1)})}$ for all $v^{(1)}\in B_1$, but
according to theorem \ref{Simulation-Fermi-1d} (\ref{even-b})
simulating $H_W^{(k v^{(1)})}$ seems not to be possible. So we
include these nearest-neighbor interactions in our initial set. The
rest of the proof is then like in the case $m=2n+1$, and we see that
the space $\mathcal{R}$ can be simulated.
\end{proof}

\section{Simulations in bosonic systems}\label{Bosonic} In the
following section we will study simulations in translationally
invariant bosonic systems using quadratic on-site Hamiltonians and
nearest-neighbor interactions. According to Eq. \eqref{L-Boson} the
generators are of the form
$$
L=\sigma^T\left(
    \begin{array}{cc}
      -W & Y \\
      -X& W \\
    \end{array}
  \right),
$$
and their commutator $L^{\prime \prime}=[L,L^{\prime}]$ is given by
\begin{eqnarray}
L^{\prime \prime}&=&{\small \left(
                    \begin{array}{cc}
                      -W^{\prime \prime} & Y^{\prime \prime} \\
                      -X^{\prime \prime} & W^{\prime \prime T} \\
                    \end{array}
                  \right)},\\
X^{\prime \prime}&=&-X^{\prime}(W+W^T)+X(W^{\prime}+W^{\prime T}),\\
Y^{\prime \prime}&=&Y(W^{\prime}+W^{\prime T})-Y^{\prime}(W+W^T),\\
W^{\prime \prime}&=&W^{\prime \prime
T}=X^{\prime}Y-XY^{\prime}.\label{W}
\end{eqnarray}
As in the fermionic case all commutators obey a symmetry which is in
this case $W''=W''^T$ corresponding to reflection symmetry (point
symmetry) of the Hamiltonian and we will denote the vector space of
all point symmetric Hamiltonians by $\mathcal{P}$. This means that
all simulated interactions are point symmetric up to linear
combinations of the initial Hamiltonians.

Every generator $L$ of an arbitrary on-site Hamiltonian is of the
form
\begin{equation}\label{Exyw}
E_{(x_E,y_E,w_E)}=\left(
            \begin{array}{cc}
              -w_E\one & y_E\one \\
              -x_E\one & w_E\one \\
            \end{array}
          \right), \;x_E, y_E, w_E \in \mathbb{R}.
\end{equation}
Generators corresponding to a nearest-neighbor interaction along
an axis $e_i$ are of the form
\begin{eqnarray}\label{HNN-Bose}
&&L^{(e_i)}=\left(
         \begin{array}{cc}
           -W & Y \\
           -X & W^T \\
         \end{array}
       \right)=\\
       &&\left(
                 \begin{array}{cc}
                 -w M^{(e_i)}+\tilde{w}M^{(-e_i)} &  y M^{(e_i)}_+\\
                  x M^{(e_i)}_+  & \tilde{w} M^{(e_i)}+w M^{(-e_i)} \\
                 \end{array}
               \right),\nonumber
\end{eqnarray}
where $x, y, w, \tilde{w} \in \mathbb{R}$ and $M^{(v)}$ has been defined in \eqref{M}.\\
 We  define $L_W^{(v)}=\big(-M_+^{(v)}\big)\oplus M_+^{(v)}
 $ and
\begin{equation}
L_X^{(v)}=\left(
              \begin{array}{cc}
                0 & 0 \\
                -M_+^{(v)} & 0 \\
              \end{array}
            \right),\; L_Y^{(v)}=\left(
              \begin{array}{cc}
                0 & M_+^{(v)} \\
                0 & 0 \\
              \end{array}
            \right),
\end{equation}
where the indices $X,Y$ and $W$ correspond to a non-zero $X$-, $Y$-
and W-block respectively.
\subsection{Simulations in one-dimensional bosonic systems} In
this section we show that for one-dimensional bosonic systems an
arbitrary nearest-neighbor interaction is complete for simulating
the vector space $\mathcal{P}$ (i.e. $W=W^T$) when supplemented by
all on-site Hamiltonians.

\begin{theorem}\label{Bose-1d}Consider bosonic systems with quadratic Hamiltonians on a one-dimensional
translationally invariant lattice with periodic boundary conditions.
The set of all possible one-mode transformations $E_{(x_E, y_E,
w_E)}$  in \eqref{Exyw} together with one arbitrary nearest-neighbor
interaction given by $L$ in Eq.\eqref{HNN-Bose} is universal for
simulating the space $\mathcal{P}$ of all point symmetric
interactions.
\end{theorem}
\begin{proof}
First we will show
that an arbitrary interaction with $X=Y=0, W=W^T$ can be brought
from the $W$-block in the $X$ and $Y$-block:
\begin{eqnarray}\label{WinXY}
\left[\left(
  \begin{array}{cc}
    -W & 0 \\
    0 & W \\
  \end{array}
\right),\left(
  \begin{array}{cc}
    0 & -\frac{1}{2} \\
    0 & 0 \\
  \end{array}
\right)\right]&=&\left(
                 \begin{array}{cc}
                   0 & W \\
                   0 & 0 \\
                 \end{array}
               \right),\\
               \left[\left(
  \begin{array}{cc}
    -W & 0 \\
    0 & W \\
  \end{array}
\right),\left(
  \begin{array}{cc}
    0 & 0 \\
    -\frac{1}{2} & 0 \\
  \end{array}
\right)\right]&=&\left(
                 \begin{array}{cc}
                   0 & 0 \\
                   -W & 0 \\
                 \end{array}
               \right).
\end{eqnarray}
Thus it is sufficient to show that an arbitrary $W$-block
 can be obtained. Let us start with a
nearest-neighbor interaction of the form $L_Y^{(e_1)}$. As
$$[L_Y^{(e_1)},E_{(1,0,0)}]=L_W^{(e_1)}$$
we also get $L_X^{(e_1)}$ according to \eqref{WinXY}. Now
$[L_Y^{(e_1)}, L_X^{(e_1)}]-2E_{(0,0,1)}=L_W^{(2e_1)},$ so that we
also get $L_X^{(2e_1)}$ and $L_Y^{(2e_1)}$. As
$[L_Y^{(ke_1)},L_X^{(e_1)}]=L_W^{((k+1)e_1)}+L_W^{((k-1)e_1)}$
we can simulate $\mathcal{P}$.\\
Finally it remains to show that we can get $L_Y^{(e_1)}$ from an
arbitrary nearest-neighbor interaction. If $y \neq 0$ in
\eqref{HNN-Bose}, then
$[[L^{(e_1)},E_{(0,0,1/2)}],E_{(-1,0,0)}]=yL_W^{(e_1)}$ so that we
get $L_Y^{(e_1)}$ according to \eqref{WinXY}. If $y=0, x \neq 0$,
then $[[L^{(e_1)}, E_{(0,0,1/2)}],E_{0,1,0}]=xL_W^{(e_1)}$ and we
get $L_Y^{(e_1)}$ as before. If $x=y=0, w \neq 0$, then $[L^{(e_1)},
E_{(0,-1,0)}]=(w+\tilde{w})L_Y^{(e_1)}.$
\end{proof}

\subsection{Simulations in $d$-dimensional bosonic systems} The
following generalizes the previous result to cubic lattices in
arbitrary spatial dimensions in cases where nearest neighbor
interactions along all axes and diagonals are available.

\begin{theorem}\label{Bose-dd-basis}
Consider a system of bosonic modes on a $d$-dimensional
translationally invariant lattice with periodic boundary conditions.
The set of all on-site transformations together with all
nearest-neighbor interactions corresponding to $L^{(\sum_i c_i
e_i)}, c_i \in\{0,\pm 1\}, i=1 \ldots, d$ with $L$ as in
Eq.(\ref{HNN-Bose})
is complete for simulating the space $\mathcal{P}$ of all possible
point symmetric interactions.
\end{theorem}

\begin{proof}
Like in the $d$-dimensional fermionic case let
$B_z=\{v|\|v\|_{\infty}\leq z \}, z\in \mathbb{N}, v^{(z)} \in B_z$.
>From Thm. \ref{Bose-1d} we know that it is sufficient to show that
$L_W^{(v)}$ can be simulated for arbitrary $v$. By induction over
$z$ we will show that $L_W^{(v^{(z)})}$ can be simulated for all
$v^{(z)} \in B_z$. For $z=1$ we know from Thm. \ref{Bose-1d} that
all interactions described by $v^{(1)}\in B_1$ can be simulated as
we have chosen our initial Hamiltonians appropriately. Now assume
that $L_W^{(v^{(z)})}$ can be simulated for all $v^{(z)}\in B_z$,
and let $v^{(z+1)} \in B_{z+1}$. Then there exist $p^{(z)}, q^{(z)}
\in B_z$ such that $p^{(z)}+q^{(z)}=v^{(z+1)}, p^{(z)}-q^{(z)}\in
B_z$ (see figure \ref{Graphik2}). As
$$[L_X^{(p^{(z)})},L_Y^{(q^{(z)})}]=L_W^{(v^{(z+1)})}-L_W^{(p^{(z)}-q^{(z)})},$$
we can simulate $L_W^{(v^{(z+1)})}$.
\end{proof}

\section{Simulations in spin systems}\label{Spin}
In this section we will consider translationally invariant quantum
lattice systems where a $D$-dimensional Hilbert space is assigned to
each of the sites. We refer to these systems as \emph{spins}
although, of course, the described degrees of freedom do not have to
be spin-like. The main result of this section is that within the
translationally invariant setting universal sets of interactions
cannot exist. These results are based on the following Lemma
involving Casimir operators, i.e., operators which commute with
every element of the Lie algebra \cite{FultonHarris, Cornwell}:

\begin{lemma}\label{Casimir}
Consider a Lie-Algebra $\mathcal{L}$ and subalgebra
$\mathcal{L}'=[\mathcal{L,L}]$. Let $\mathcal{S_L}$ be a set of
generators for $\mathcal{L}$ and $C$  a Casimir operator of
$\mathcal{L}$ fulfilling
\begin{equation}\label{TR}
\mbox{tr}[CG]=0\; \forall G \in
\mathcal{S_L}\setminus\mathcal{L}'.
\end{equation}
Then for every $K \in \mathcal{L}$ we have that $\mbox{tr}[CK]=0$.
\end{lemma}

\begin{proof}
Every $K \in \mathcal{L}$ can be written as
\begin{equation}\label{Casi}
K=\sum_{L\in \mathcal{L}'} \alpha_L L + \sum_{G\in
\mathcal{S_L}\setminus\mathcal{L}'} \beta_G G\;,\ \
\alpha_L,\beta_G\in\mathbb{C}.
\end{equation}
Since we can write any $L\in\mathcal{L}'$ as $L=[L_1,L_2]$,
$L_i\in\mathcal{L}$ we have that
$$\mbox{tr}[CL]=\mbox{tr}[C[L_1,L_2]]=\mbox{tr}[[CL_1,L_2]]=0$$
where we have used that $C$ is a Casimir operator, i.e., $\forall
L\in\mathcal{L}: [C,L]=0$. Hence if we take the trace of
Eq.(\ref{Casi}) with $C$ we get
$$\mbox{tr}[CK]=\sum_{G \in
\mathcal{S_L}\setminus\mathcal{L}'} \beta_G \mbox{tr}[CG]$$ which
vanishes according to the assumption in Eq.(\ref{TR}).
\end{proof}
Let us now exploit Lemma \ref{Casimir} in the translationally
invariant setting in order to rule out the universality of
interactions corresponding to certain sets of generators
$\mathcal{S_L}$. The following results are stated for
one-dimensional systems but they can be applied to $d$-dimensional
lattices by grouping sites in $d-1$ spatial dimensions.

We use Casimir operators of the form
\begin{equation}
\label{Cop} C=\sum_{k=1}^m \gamma_k
T^k\;,\quad\gamma_k\in\mathbb{C}\;,
\end{equation}
where $T|i_1,i_2,\ldots,
i_m\rangle=|i_2,i_3,\ldots,i_m,i_1\rangle$ is the translation
operator which shifts the lattice by one site. To simplify
notation we define an operator
\begin{equation}
\tau(X)=\sum_{j=1}^m T^j X T^{\dagger j}\;,
\end{equation} which symmetrizes any operator $X$ with respect to the translation group. If $X$ does not act on the entire lattice
we will slightly abuse notation and write $X$ instead of
$X\otimes\one\otimes\cdots\otimes\one$.

\begin{theorem}\label{factorthm}
Consider a translationally invariant spin system on a ring of length
$m$. If $f$ is a non-trivial factor of $m$ then there is no
universal set of Hamiltonians with interaction range smaller than
$f$ which generates all translationally invariant interactions. In
particular if $m$ is even, nearest-neighbor interactions cannot
generate all next-to-nearest neighbor Hamiltonians.
\end{theorem}
\begin{proof}
Let us introduce a basis of the Lie algebra $su_{D^m}$ of the form
$i \sigma_{j_1}\otimes\cdots\otimes\sigma_{j_m}$,
$j_k\in\mathbb{Z}_{D^2}$ where $\sigma_k$ is traceless except for
$\sigma_0=\one$ and $\mbox{tr}[\sigma_k^2]=2$ for all $k>0$. For
$D=2$ these are the Pauli matrices
\begin{equation}\label{Pauli}
\sigma_1=\left(
                                \begin{array}{cc}
                                  0 & 1 \\
                                  1 & 0 \\
                                \end{array}
                              \right),\; \sigma_2=\left(
                                                   \begin{array}{cc}
                                                     0 & -i \\
                                                     i & 0 \\
                                                   \end{array}
                                                 \right),\;\sigma_3=\left(
                                                                     \begin{array}{cc}
                                                                       1 & 0 \\
                                                                       0 & -1 \\
                                                                     \end{array}
                                                                   \right),
\end{equation}
 and for $D>2$ we can simply choose all possible embeddings
 thereof. Let $m=f\cdot f'$ and consider a Casimir operator of the
 form in Eq.(\ref{Cop}) with $\gamma_k=\delta_{k,f}$, i.e., $C=T^f$.

 We first show that for every Hamiltonian $H\in
 su_{D^m}$ with interaction range smaller than $f$ we have
 $\mbox{tr}[CH]=0$. To see this note that the shift operator $T^f$ contracts the trace of a tensor
 product as
 \begin{equation}\label{zerotrace}
 \mbox{tr}\big[T^f \sigma_{j_1}\otimes\cdots\otimes\sigma_{j_m}
 \big]= \prod_{\beta=1}^f \mbox{tr}\left[\prod_{\alpha=0}^{f'-1}\sigma_{j_{(\beta+\alpha
 f)}}\right].
 \end{equation}
 In order to arrive at formula \eqref{zerotrace} expand the
 translation operator $T^f$ in the computational basis
 $$ T^f= \sum_{i_1, \ldots, i_m} | i_1 \ldots i_m \rangle \langle i_{m-(f-1)} i_{m-(f-2)} \ldots i_1 \ldots i_{m-f}|$$
in order to get
\begin{eqnarray*}
\mbox{tr}\big[T^f \sigma_{j_1}\otimes\cdots\otimes\sigma_{j_m}
 \big]=\hspace{5cm}\\
\sum_{i_1, \ldots, i_m}\langle i_{m-(f-1)}|\sigma_{j_1}|i_1 \rangle
\langle i_{m-(f-2)}|\sigma_{j_2}|i_2 \rangle \ldots
 \langle i_{m-f}|
\sigma_{j_m} | i_m \rangle.
\end{eqnarray*}
Rearranging the order of the factors and using $(f^{\prime}-1)f=m-f$
leads to
\begin{eqnarray*}
\prod_{\beta =1}^f\sum_{i_\beta, i_{f+\beta}, \ldots}\langle
i_{m-(f-\beta)}|\sigma_{j_{\beta}}|i_{\beta} \rangle
\langle i_{\beta}|\sigma_{j_{(f+\beta)}}|i_{f+\beta} \rangle \cdot \ldots \cdot \\
 \langle i_{(f^{\prime}-2)f+\beta}|
\sigma_{j_{((f^{\prime}-1)f+\beta)}} | i_{(f^{\prime}-1)f+\beta}
\rangle=\\
\prod_{\beta=1}^f
\mbox{tr}\left[\prod_{\alpha=0}^{f'-1}\sigma_{j_{(\beta+\alpha
 f)}}\right]\hspace{4.5cm}
\end{eqnarray*}

 If the interaction range of $H$ is smaller than $f$ then it can
 be decomposed into elements of the form
 $\tau(\sigma_{i_1}\otimes\cdots\otimes\sigma_{i_f})$. Since $\sigma_k$ is traceless for all $k>0$, all these terms amount to a vanishing trace in
 Eq.(\ref{zerotrace})
 so that we have indeed $\mbox{tr}[CH]=0$. This means we can apply Lemma \ref{Casimir}
 to the set $\mathcal{S_L}$ corresponding to all interactions with range smaller than $f$.

 Now consider a two-body interaction between site one and
 site $f+1$ of the form
 $\tilde{H}=\tau\big( \sigma_j^{(1)}\sigma_j^{(1+f)}\big)$. From Eq.(\ref{zerotrace}) we get
 $$\mbox{tr}[C\tilde{H}]=m\mbox{tr}[\sigma_j^{(1)}\sigma_j^{(1+f)}]\prod_{\beta=2}^f\mbox{tr}[\one]=2m D^f$$
 such that by Lemma \ref{Casimir} we conclude that $\tilde{H}$ cannot be simulated.
\end{proof}

The following shows that a universal set of nearest-neighbor
interactions cannot exist irrespective of the factors of $m$:
\begin{theorem}
Consider a ring of length $m$. Then the set $\mathcal{S_L}$
corresponding to all on-site Hamiltonians and nearest-neighbor
interactions is not universal for simulating all translationally
invariant Hamiltonians. In particular for $D=2$ a product
Hamiltonian
\begin{equation} H=\tau
\Big(\sigma_{j_1}\otimes\cdots\otimes\sigma_{j_m}\Big)
\end{equation}
cannot be simulated if  $\sigma_1,\sigma_2$ and $\sigma_3$ all
occur an odd number of times.
\end{theorem}
\begin{proof}
We use the Casimir operator $C=T-T^\dagger$ and the set of
generators $\mathcal{S_L}=i\{\tau(\sigma_k\otimes\sigma_l)\}\subset
su_{2^m}$. As $\mbox{tr}[CG]=0$ for all $G\in\mathcal{S_L}$  we can
again apply Lemma \ref{Casimir}. Consider now the above product
Hamiltonian $H$ or if $D>2$ its embedding respectively. Using
Eq.(\ref{zerotrace}) with $f=1, f^{\prime}=m$ we obtain
\begin{equation}
\mbox{tr}[CH]=m\mbox{tr}\left[\prod_{k=1}^m
\sigma_{i_k}-\prod_{l=1}^m\sigma_{i_l}^T \right].
\end{equation}
Since $\sigma_i^T=(-1)^{\delta_{i,2}}\sigma_i$ and by assumption
$\sigma_2$ appears an odd number of times we get $\mbox{tr}[CH]=2
m\mbox{tr}\big[\prod_{k=1}^m \sigma_{i_k}\big]$ which is non-zero
iff $\sigma_1$ and $\sigma_3$ appear and odd number of times as
well.
\end{proof}

Clearly, one can derive other no-go theorems in a similar manner
from Lemma \ref{Casimir}. However, we end this section by
providing some examples of interactions which \emph{can} be
simulated. For this we define
$g_{kl}=\tau(\sigma_k\otimes\sigma_l)$.

\begin{theorem}
Consider a translationally invariant system of $m$ qubits ($D=2$)
 on a ring. By using on-site Hamiltonians and nearest-neighbor interactions the following interactions can be simulated:
\begin{eqnarray}
&&\tau(\sigma_i \otimes \sigma_i \otimes \sigma_i),\label{3sigma}\\
J_{ij}^{(r_j)}&=&\tau(\sigma_i \otimes \sigma_j \otimes \ldots
\sigma_j \otimes \sigma_i)\label{sigma-fermi},
\end{eqnarray}
where $i,j \in \{1,2,3\}$ and $r_j$ denotes the number of $\sigma_j$
matrices. Moreover, for $m=5$ one can simulate next-to-nearest
neighbor interactions of the form $N_i=\tau(\sigma_i \otimes \one
\otimes
  \sigma_i\otimes \one \otimes \one).$
\end{theorem}

\begin{proof}
We will restrict our proof to the pairs $i=1, j=2$ as the other
interactions can be obtained in an analogous way.  We start proving
(\ref{3sigma}).  The Hamiltonian \eqref{3sigma} can be simulated due
to $[g_{21},g_{13}]/(2i)=\tau(\sigma_1 \otimes \sigma_1\otimes
\sigma_1)$. For proving \eqref{sigma-fermi}, we start with
$[g_{13},g_{11}]/(2i)=J_{12}^{(1)}$. As
$[J_{12}^{(r_2)},g_{31}]/(2i)=J_{12}^{(r_2-1)}-J_{12}^{(r_2+1)}$,
$J_{12}^{(0)}=g_{11}$, we have shown \eqref{sigma-fermi}.\\
Now we will prove that the next-to-nearest neighbor interaction can
be achieved for $m=5$, $i=1$ ($i=2,3$ follow similarly). Using
\eqref{sigma-fermi}, we see that $\tau(\sigma_1\otimes \sigma_1
\otimes \sigma_1\otimes \sigma_1\otimes \one)$ can be simulated, as
$[g_{23},J_{21}^{(3)}]/(2i)+J_{(21)}^{(2)}=\tau(\sigma_1\otimes
\sigma_1 \otimes \sigma_1\otimes \sigma_1\otimes\one)$, and
similarly we get $\tau(\sigma_3\otimes \sigma_3 \otimes
\sigma_3\otimes \sigma_3\otimes\one)$. As
$[[g_{11},g_{23}],g_{32}]/4=2N_1- \tau(\sigma_1\otimes \sigma_1
\otimes \sigma_1\otimes \sigma_1\otimes \one) - \tau(\sigma_3\otimes
\sigma_3 \otimes \sigma_3\otimes \sigma_3\otimes \one)$ we can
extract $N_1$.
\end{proof}

\section{Conclusions}

We have presented a characterization of universal sets of
translationally invariant Hamiltonians for the simulation of
interactions in quadratic fermionic and bosonic systems given the
ability of engineering local and nearest neighbor interactions.
Thereby the Lie algebraic techniques of quantum simulation restrict
the space of reachable interactions to Hamiltonians with real
hopping amplitudes in the case of fermions and to point symmetric
interactions in the case of bosons.

For spins the situation appears to be more difficult and a
complete characterization of interactions that can be simulated
remains to be found. As a first step, we have identified
Hamiltonians that cannot be simulated using short range
interactions only. Furthermore, we have introduced a technique
based on the Casimir operator of the corresponding Lie algebra
which allows one to find Hamiltonians that cannot be simulated
with a given set of interactions.

In this work we have considered the question of what can be
simulated leaving aside the question of the efficiency. In this
context it is important to remark  the fact that the number of
applications of the original Hamiltonians in order to obtain a
result bounded by some given error scales polynomially in the
Trotter expansion. The scaling with the total number of particles
depends on the number of commutators that are required to obtain
the Hamiltonian.

Finally, whereas we have shown that it is not possible to perform
certain simulations for spin systems it is still possible to
perform those simulations by encoding the qubits in a different
way.

We thank the Elite Network of Bavaria QCCC, DFG-Forschungsgruppe
635, SFB 631, SCALA and CONQUEST.


\begin{thebibliography}{99}


\bibitem{Feynman}
R.P. Feynman, Int. J. Theor. Phys. {\bf 21}, 467 (1982).

\bibitem{Lloyd}
S. Lloyd,
 Science {\bf 273}, 1073 (1996);
C. Zalka, Proc. Roy. Soc. Lond. A {\bf 454}, 313 (1998).

\bibitem{opticallattices} M. Greiner, O. Mandel, T. Esslinger,
T.W. H\"ansch, I. Bloch, Nature(London) {\bf 415}, 39 (2002).

\bibitem{iontraps} D. Leibfried, R. Blatt, C. Monroe, D. Wineland,
Rev.Mod. Phys. {\bf 75}, 281 (2003).

\bibitem{Diego} D. Porras, J.I. Cirac, quant-ph/0401102;
quant-ph/0409015; quant-ph/0601148.

\bibitem{Jane}  E. Jane, G. Vidal, W. D\"ur, P. Zoller, J.I.
Cirac, Quant. Inf. Comp. {\bf 3}(1), 15 (2003).

\bibitem{QPTheorie} D. Jaksch, C. Bruder, J.I. Cirac, C.W.
Gardiner, P. Zoller, Phys. Rev. Lett. {\bf 81}, 3108 (1998).

\bibitem{2universality} D. Deutsch, A. Barenco, A. Ekert, Proc. R.
Soc. Lond. A {\bf 449}, 669 (1995); S. Lloyd, Phys. Rev. Lett.,
{\bf 75}, 346 (1995).

\bibitem{2body} J.L. Dodd, M.A. Nielsen, M.J.
Bremner, R.T. Thew, Phys. Rev. A {\bf 65}, 040301(R) (2002); M.A.
Nielsen, M.J. Bremner, J.L. Dodd, A.M. Childs, C.M. Dawson, Phys.
Rev. A {\bf 66}, 022317 (2002).

\bibitem{manybody} M. J. Bremner, J. L. Dodd, M. A. Nielsen, and D. Bacon,
Phys. Rev. A {\bf 69}, 012313 (2004);  M. J. Bremner, D. Bacon, M.
A. Nielsen, Phys. Rev. A {\bf 71}, 052312 (2005).



\bibitem{eff} C.H. Bennett, J.I. Cirac, M.S. Leifer, D.W. Leung, N.
Linden, S. Popescu, G. Vidal, Phys. Rev. A {\bf 66}, 012305 (2002);
P.Wocjan, M. R\"otteler, D. Janzing, T. Beth, Phys. Rev. A {\bf 65},
042309 (2002); P. Wocjan, M. R\"otteler, D. Janzing, T. Beth,
Quantum Inf. Comput. {\bf 2} 133 (2002); G. Vidal, J.I. Cirac, Phys.
Rev A {\bf 66}, 022315 (2002).






\bibitem{FultonHarris} W. Fulton, J. Harris, Representation Theory,
Graduate Texts in Mathematics {\bf 129}, Springer (1991).

\bibitem{Cornwell} J.F. Cornwell, Group Theory in Physics Vol. 1
and 2, Academic Press.

\bibitem{LieTrotter} H.F. Trotter, Proc. Am. Math. soc. {\bf 10},
545 (1959); P.R. Chernoff, J. Functional Analysis {\bf 2}, 238
(1968).

\bibitem{BCH} F. Haussdorff, Ber. Ver. Saechs. Akad. Wiss. Leipzig,
Math.-Phys.Kl. {\bf 58}, 19 (1906); E.B. Dynkin Math. Rev. {\bf 11},
80 (1949); E.B. Dynkin, Am. Math. Soc. Transl. {\bf 9}, 470 (1950).

\bibitem{LSM} E. Lieb, T. Schulz, D. Mattis, Ann. Phys. (N.Y.) {\bf
16}, 407 (1961).

\bibitem{Araki} H. Araki, "Bogoliubov tranformation and Fock
representation of canonical anticommutator relations" in Operator
Algebras and Mathematical Physics, Contemporary Mathematics {\bf
62}, Iowa City (1987).

\bibitem{Bravyi} S. Bravyi, Quantum Inf. and Comp. {\bf 5}, 216
(2005).

\bibitem{Wolf} M.M. Wolf, Phys. Rev. Lett. {\bf 96}, 010404 (2006).

\bibitem{Schuch}N. Schuch, J.I. Cirac, M. Wolf, Commun. Math. Phys. {\bf 267}, 65
(2006).

\bibitem{Audenaert-Eisert-Werner}  K. Audenaert, J. Eisert, M.B. Plenio, R.F. Werner,
Phys. Rev. A {\bf 66}, 042327 (2002).


\end{thebibliography}
\end{document}